\documentclass[conference]{IEEEtran}
\IEEEoverridecommandlockouts
\usepackage{booktabs}
\usepackage{cite}
\usepackage{amsmath,amssymb,amsfonts}
\usepackage{algorithmic}
\usepackage[ruled,vlined]{algorithm2e}
\usepackage{graphicx}
\usepackage{textcomp}
\usepackage{xcolor}
\def\BibTeX{{\rm B\kern-.05em{\sc i\kern-.025em b}\kern-.08em
    T\kern-.1667em\lower.7ex\hbox{E}\kern-.125emX}}
\begin{document}

\title{Open RAN-Enabled Deep Learning-Assisted Mobility Management for Connected Vehicles}

\author{\IEEEauthorblockN{1\textsuperscript{st} Maria Santana}
\IEEEauthorblockA{\textit{Centro de Informática (CIn)} \\
\textit{Universidade Federal de Pernambuco (UFPE)}\\
Recife, Brasil \\
mksb@cin.ufpe.br}
\and
\IEEEauthorblockN{2\textsuperscript{nd} Kelvin Lopes Dias}
\IEEEauthorblockA{\textit{Centro de Informática (CIn)} \\
\textit{Universidade Federal de Pernambuco (UFPE)}\\
Recife, Brasil \\
kld@cin.ufpe.br}
}

\maketitle

\begin{abstract}
Connected Vehicles (CVs) can leverage the unique features of 5G and future 6G/NextG networks to enhance Intelligent Transportation System (ITS) services. However, even with advancements in cellular network generations, CV applications may experience communication interruptions in high-mobility scenarios due to frequent changes of serving base station, also known as handovers (HOs). This paper proposes the adoption of Open Radio Access Network (Open RAN/O-RAN) and deep learning models for decision-making to prevent Quality of Service (QoS) degradation due to HOs and to ensure the timely connectivity needed for CV services. The solution utilizes the O-RAN Software Community (OSC), an open-source O-RAN platform developed by the collaboration between the O-RAN Alliance and Linux Foundation, to develop xApps that are executed in the near-Real-Time RIC of OSC. To demonstrate the proposal's effectiveness, an integrated framework combining the OMNeT++ simulator and OSC was created. Evaluations used real-world datasets in urban application scenarios, such as video streaming transmission and over-the-air (OTA) updates. Results indicate that the proposal achieved superior performance and reduced latency compared to the standard 3GPP HO procedure.

\end{abstract}


\section{Introduction}
Vehicular communications have gained prominence with the advent of fifth-generation (5G) systems and are expected to be further enhanced by the 6G/NextG networks envisioned for 2030 \cite{6g}. With a global market size projected to reach around USD 181.90 billion by 2034 \cite{car_market}, connected cars are an essential part of Intelligent Transportation Systems (ITS) and can benefit from various applications and services \cite{b2} provided by the wireless connectivity of current and future 3rd Generation Partnership Project (3GPP) mobile network infrastructures. Potential vehicular applications enabled by 3GPP networks can range from video streaming for infotainment or autonomous driving assistance, to over-the-air (OTA) updates.

On one hand, this ecosystem is complex, as each aforementioned application requires specific network performance indicators and generates different types and volumes of data. On the other hand, maintaining uninterrupted connectivity in vehicular networks remains a challenge, especially due to car’s high mobility, which affects applications’ quality of service (QoS) due to frequent handovers, i.e., changing of base station during ongoing sessions. The handover procedure, when not seamlessly executed, can cause delays, ping pong effects, signal degradation, and unnecessary changes of Next Generation Node B (gNodeB). This impacts directly the QoS of the user, such as increased loss rate and latency \cite{b4}. Thus, vehicular networks will require flexible, adaptive, and intelligent frameworks to deal with the intrinsic features of mobility, wireless connectivity issues, and vehicular applications requirements to provide the service continuity and expected outcomes for the whole ITS.

Recently, the Open RAN (O-RAN) \cite{oran} architecture has emerged to promote open interfaces, programmability, and intelligence at the radio access network (RAN) level. This approach aims to avoid vendor lock-in and foster flexibility and innovation within the critical wireless access infrastructure of cellular networks. O-RAN builds on the disaggregation of the 3GPP-defined, previously monolithic base station components—namely, the Radio Unit (RU), Distributed Unit (DU), and Centralized Unit (CU)—by establishing open interfaces between these elements. Consequently, RAN components from different vendors can interoperate seamlessly thanks to O-RAN’s open architecture.

Central to the concept of O-RAN is the software-defined RAN Intelligent Controller (RIC), which operates from two different control loop perspectives: the near-Real-Time RIC (near-RT-RIC), functioning at intervals between 10 ms and 1 s, and the Non-RT RIC, which operates on timescales longer than one second. While the latter oversees long-term RAN management and optimization, as well as the training of machine learning (ML) and artificial intelligence (AI) models, the former operates on a finer-grained timescale, handling functions such as handover, load balancing, RAN slicing, and other tasks that require execution in less than 1 s.

It is worth noting that the near-RT RIC communicates with gNodeB, collects data, and exposes control primitives through the standardized O-RAN E2 interface \cite{b5}. On the other hand, communication between the two RICs occurs through the A1 interface. Network applications, known as xApps and rApps, are executed on top of the near-RT RIC and the non-RT RIC, respectively. These applications run in a virtualized environment as microservices.

The O-RAN Alliance has included vehicular communications as one of their use case scenarios \cite{user_case_v2x}. This integration has been discussed and proposed in several recent articles. In \cite{b6}, the authors model an O-RAN-driven millimeter-wave beam management system for a cooperative handover approach using mathematical regression and developed their own high-level Python simulator. However, \cite{b6} neither implements an xApp nor employs an RIC or 5G simulation platform to evaluate the proposal. A machine learning-based model to predict compatibility time—i.e., the period during which a vehicle remains within the communication range of another—was proposed in \cite{b7}. \cite{b20} proposed a xApp to select relay nodes in millimeter waves. To validate their proposal, the authors used ns-3 alongside the Open RAN framework, however, the current 5G ns-3 modules lack handover functionality. In \cite{b21}, Open RAN is proposed for traffic steering, prioritizing platoons and first responders, but without handover optimization or an open-source solution.

To the best of our knowledge, current state-of-the-art Open RAN-based solutions for vehicular networks do not address HOs among gNodeBs in a true 5G simulation environment using deep learning models to meet user QoS requirements for applications such as video streaming and OTA updates. So, the aim of this paper is twofold: First, it proposes a new joint 5G simulation and O-RAN Software Community (OSC) RIC/development framework for evaluating proposals and the synergy between O-RAN and vehicular communications. Secondly, the paper demonstrates the effectiveness of the framework and addresses infrastructure access and handover (HO) issues for connected cars in 5G and beyond by proposing an O-RAN-based HO solution using deep learning models.

\begin{itemize}

    \item \textbf{An integrated fully open-source platform composed of OSC and the OMNeT++ simulator:} We have developed an environment for O-RAN that can potentially connect large-scale 5G simulations in OMNeT++ and leverage the near-RT RIC of OSC to run xApps. We extended the OMNeT++’s 5G library by incorporating a module with an E2 interface, facilitating the development and testing of xApps before deploying them in real-world operational scenarios. The relevant system design details are discussed in Section \ref{sec:arquitetura}.
    \item \textbf{Access and Handover Decision:} We developed a deep learning-based data-driven HO xApp using the KPM Monitor xApp. We employed OMNeT++/O-RAN to collect data, design, and test the HO xApp. The proposal is detailed in Section \ref{sec:proposal}.
    \item \textbf{Performance Evaluation:} We evaluated the xApp using a mobility scenario based on a real dataset and different use cases: MEC-based video streaming and OTA updates. Performance evaluation metrics include Channel Quality Indicator (CQI), Throughput, Delay and Session Freeze. Section \ref{sec:simulation} presents the evaluation and analysis;
\end{itemize}

\section{Integrated OSC and OMNeT++ Simulator Evaluation Framework}
\label{sec:arquitetura}

Using realistic channel and protocol stack models enables data collection for machine learning-based xApps, reducing the need for costly, large-scale real-world deployments. Figure \ref{fig:arquitecture} presents the proposed framework, emphasizing the interaction between the Near-RT RIC and a gNodeB.

\subsection{O-RAN Software Community and its components}

The O-RAN Software Community (OSC), an open-source initiative by the O-RAN Alliance and Linux Foundation, manages software development in line with O-RAN Alliance specifications. Their efforts include components for both Near-RT and Non-RT RIC, along with xApp examples, like Traffic Steering and Anomaly Detection (AD), as well as simulators (e2sim). The Non-RT RIC, which operates as part of the Service Management and Orchestration (SMO) framework, is responsible for functionalities such as providing targeting and enrichment information via policies and managing machine learning models, including generating and updating them.

\begin{figure*}[btp]
    \centering
    \includegraphics[width=0.68\linewidth]{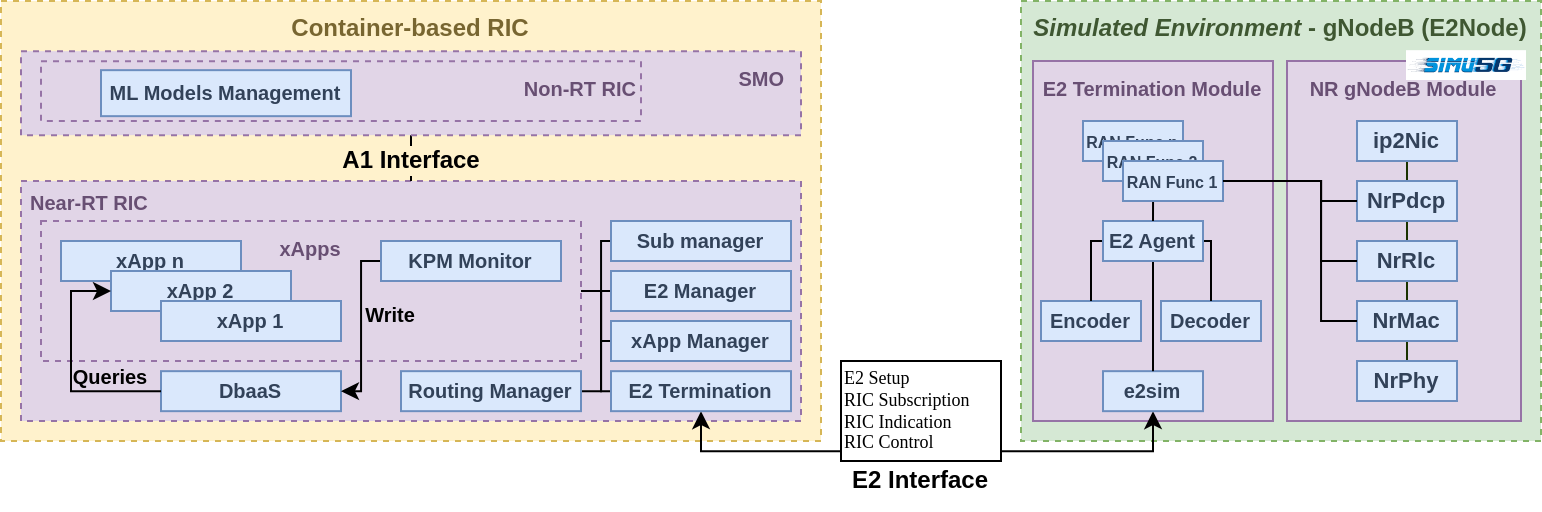}
    \caption{Integrated O-RAN Software Community and Simu5G Framework.}
    \label{fig:arquitecture}
\end{figure*}

The main functionalities of the Near-RT RIC involve the management of E2Nodes, xApps, and the interactions between them. The E2 interface connects the Near-RT RIC to the E2 nodes. As illustrated in Figure \ref{fig:arquitecture}, this interface enables the execution of different control cycles, originating a sequence of procedures. Among the procedures implemented in this work are: E2 Setup, which is responsible for creating the SCTP communication and registering the list of RAN functionalities (RAN Func) available in an E2Node, such as cell and user monitoring, in addition to handover management. The RIC Subscription handles the subscription of xApps for specific functionality in an E2Node since each xApp is a microservice responsible for a specific task, such as the Key Performance Metrics (KPM) Monitor, which receives the metrics from the E2Node and stores them in the database. Thus, for network optimization and control, the use of several xApps becomes necessary. The RIC Indication sends notifications from the gNodeB to the RIC, and the RIC Control allows the RIC to send control actions, originated by the xApps, to the gNodeB.

To support the xApps and allow these procedures to be carried out, some RIC components are necessary, such as the xApp Manager, which manages the xApps lifecycle, including deployment and coordination; the Subscription Manager, which manages the xApps subscriptions for the E2Nodes available in the RIC; the E2 Manager, responsible for handling the E2 interface and facilitating communication between the Near-RT RIC and the E2Nodes; the E2 Termination, which manages the signaling between the RIC and the gNodeB through the E2 interface using the SCTP protocol; the Routing Manager, responsible for routing messages between different RIC components; and the Database-as-a-Service (DBaaS), using Redis as the underlying database, stores the data collected by various xApps and makes it accessible through the Shared Data Layer (SDL) API for queries and analysis. Integration and testing with OMNeT++ in this paper used Near-RT provided by OSC, more specifically Release I.

\subsection{OMNeT++ Simulator}

OMNeT++ provides modules to build simulation environments for both wired and wireless networks using C++ language  \cite{omnetpp}. Simu5G is one of the modules in OMNeT++, designed to simulate data plane communications in 5G. The RAN implemented in Simu5G has protocol layers 3GPP-compliant. The NR module consists of the following components: the ip2Nic, which manages packet processing and interfaces; the Packet Data Convergence Protocol (NrPdcp), responsible for header compression, encryption, and data segmentation; the Radio Link Control (NrRlc), which handles error correction and retransmission for the data link layer; the Medium Access Control (NrMac), which manages data scheduling and resource allocation; and the Physical Layer (NrPhy), responsible for the physical transmission of radio signals \cite{b9}.

The architecture of Simu5G provides a modular and flexible framework that allows the Near-RT RIC to dynamically control and monitor gNodeBs through xApps and the E2 interface. To enable this, an E2 Termination Module with an E2 Agent has been developed to manage communication with the RIC, bridging the RIC and RAN function callbacks. The RAN Functions encompass specific functionalities, such as handover management and user KPI monitoring, with which the RIC interacts through callbacks. These callbacks receive the E2AP PDU and, in turn, communicate with the Simu5G radio protocol stack to execute the requested actions, such as sending RIC Indications with user information to the KPM Monitor or triggering handover events. The Encode/Decode component is responsible for encoding and decoding ASN.1 messages exchanged between the RIC and the gNodeB over the E2 interface. Finally, e2sim \cite{e2sim} integrates the E2 Application Protocol (E2AP), managing the actual protocol for the E2 interface, and ensuring that communication between the RIC and the gNodeB complies with O-RAN standards.

\section{O-RAN-Based Handover Optimization}
\label{sec:proposal}

This section presents the architectural components and the handover mechanism in vehicular networks proposed in this paper. The proposed solution aims to optimize handover, using deep learning and incorporating data captured by the KPM monitor xApp.The architecture presented in Figure \ref{fig:arquitecture} consists of four main blocks: (a) The KPM monitor xApp (KPM Mon.) is responsible for capturing user information, and storage on the database. (b) The handover management xApp (HO Mgmt) is responsible for monitoring users and verifying the need for handover. (c) The Quality of Service Predictor (QP) xApp is responsible for generating predictions requested by the HO Mgmt, enabling proactive anticipation of handover requirements when necessary, and (d) the module responsible for training and generating the deep learning models, which uses the available database and can offload the trained model to the QP xApp;

\begin{figure}[ht!]
    \centering
    \includegraphics[width=0.60\linewidth]{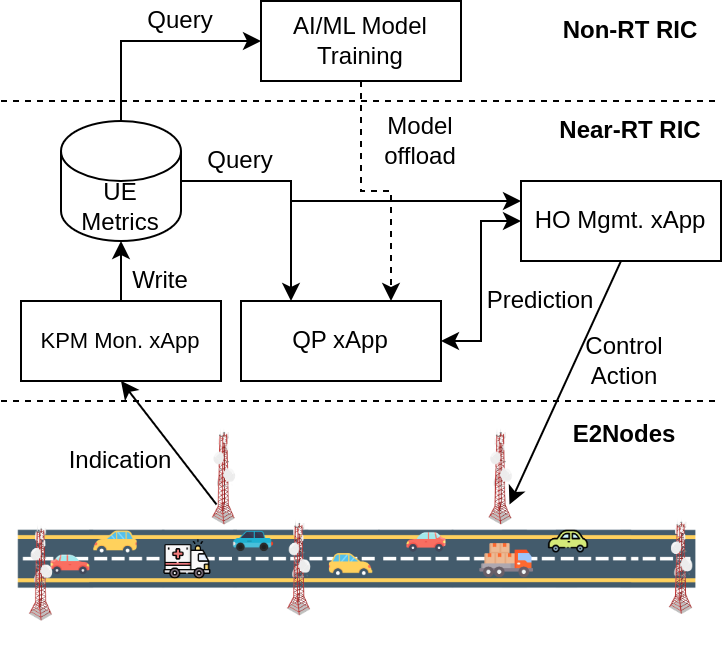}
    \caption{xApps and its Interactions in the proposed HO solution.}
    \label{fig:proposal}
\end{figure}

\subsection{Handover Mechanism}

\begin{figure}[ht!]
    \centering
    \includegraphics[width=1\linewidth]{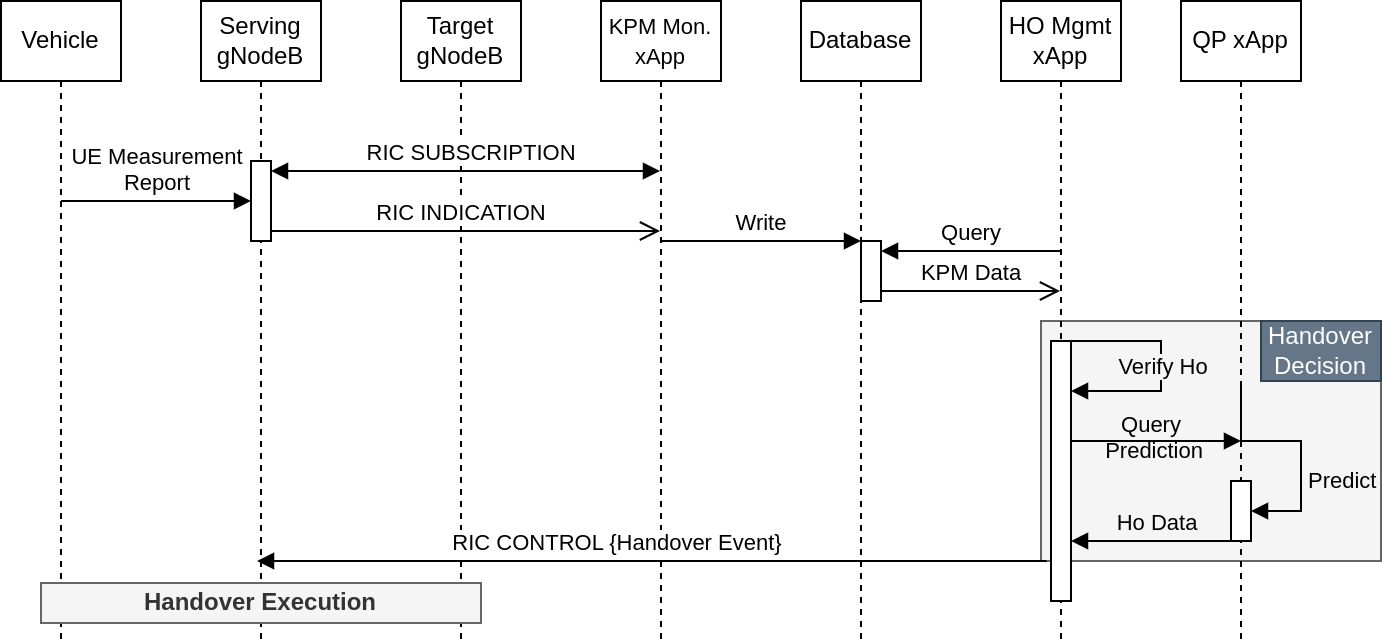}
    \caption{Signaling Flow of the Proposed Solution.}
    \label{fig:signaling}
\end{figure}

This paper proposes optimizing the handover procedure using an Open RAN-based approach, grounded on real-time data, to feed a deep learning model aimed at predicting signal quality. Thus, this section describes the proposed solution's logical operation and signaling flow, presented on the Figure \ref{fig:proposal} and \ref{fig:signaling}. Initially, the KPM Monitor submits a subscription request to the monitoring service of an E2 Node, specifying the desired metrics. The E2 Node associates this request with a callback, fulfilling the RIC subscription request. Subsequently, the vehicle sends measurement reports to the gNodeB, which converts this information into a RIC Indication to be sent to the KPM Monitor. This data is transmitted to the xApp every second. Upon receiving the metrics, the xApp stores them in a database, making them accessible for other xApps in the decision-making process. 

Once the user metrics are available in the database, the HO Mgmt periodically queries this information to assess the need for handover predictions. This verification occurs as follows: when the Reference Signal Received Power (RSRP) of a neighboring gNodeB, combined with the Handover Hysteresis Margin (HOM), exceeds that of the serving cell $(RSRP_{tgt} + HOM > RSRP_{serv})$, the QP xApp receives a call requesting a signal prediction for a specific user. The QP xApp performs \textit{N} predictions of future signals using a deep learning model. During the prediction process, the algorithm first checks the inversion of RSRPs $(RSRP_{tgt} > RSRP_{serv})$; if it occurs, the algorithm converts the prediction iteration into a Time to Trigger (TTT). The prediction continues until Event A3 occurs $(RSRP_{tgt} - HOM > RSRP_{serv})$, ensuring the complete transfer of the user to the neighboring cell or until the \textit{N} interactions finish. If the condition for Event A3 is met, the serving cell receives the stored TTT via a RIC Control Action message containing the \textit{Handover Event}, and subsequently performs the handover execution.

\subsection{Implementation of Supervised prediction models}

This section describes the methodology employed in designing the ML models, including hyperparameter selection, number of layers, and number of neurons. The chosen models were the Gated Recurrent Unit (GRU) and Long Short-Term Memory (LSTM), both widely recognized in sequential data processing tasks. For these models, defining the number of layers and units is crucial, as it directly affects performance. Since no fixed rule exists, we conducted experiments to identify an efficient configuration while preventing overfitting. Table \ref{tab:SuperLearn} presents the selected values.

\begin{table}[h!]
 \caption{Description of Implemented Models}
\centering
 \begin{tabular}{c c c c} 
 \hline
 \textbf{Model} & \textbf{Hidden Layer} & \textbf{Units} & \textbf{Prediction} \\ [0.5ex] 
 \hline
 \textbf{LSTM} & 2 LSTM, 2 Dropout & 64, 32, 0.2 & Signal \\
 \textbf{GRU} & 1 GRU & 128 & Signal  \\ [1ex] 
 \hline
 \end{tabular}
\label{tab:SuperLearn}
\end{table}
 
For hyperparameter optimization, we employed Grid Search. The models underwent training and cross-validation to find the optimal hyperparameter combination that produced the best performance. We selected the values yielding the lowest Mean Absolute Error (MAE) and Mean Squared Error (MSE) during training and validation. Table \ref{table:parameters} summarizes the tested hyperparameter sets.
 

\begin{table}[ht]
 \caption{Values of tested hyperparameters}
\centering
 \begin{tabular}{c c} 
 \hline
\multicolumn{2}{ c }{\textbf{Hyperparameters}}\\
 \hline
    Look Back & 10 and 15 \\
    Optimizer & RMSProp and Adam \\
    Activation Function & Relu and Linear\\
    Batch Size & 16, 32 and 64 \\ 
    Learning Rate & 0.0001, 0.0005, 0.001 and 0.005 \\[1ex] 
 \hline
 \end{tabular}
\label{table:parameters}
\end{table}

Considering these hyperparameters, we initiated the training phase using the dataset provided in \cite{b10}. This dataset includes various 5G network metrics relevant to vehicular networks, such as RSRP, Reference Signal Received Quality (RSRQ), Signal-to-Noise Ratio (SNR), Bitrate, jitter and packet loss. After completing the training and validation phases, the selected hyperparameters for both the GRU and LSTM models were as follows: a batch size of 16, a learning rate of 0.0001, a lookback period of 15, and the ReLU activation function. The GRU model utilized the Adam optimizer, while the LSTM model employed RMSProp.

\section{Use-Cases Performance Evaluation}
\label{sec:simulation}

This evaluation aims to analyze the performance of the proposed solution for vehicular networks within a 5G Open RAN architecture. The framework focuses on enhancing QoS by improving handover efficiency. We seek to determine whether the solution improves network performance indicators across various use cases. This section first describes the baseline handover mode for comparison, the simulation scenarios, and the relevant metrics. We then discuss the results from evaluations conducted across different use cases.

\subsection{Baseline}
The model used as the baseline was the 3GPP standard and included in the Simu5G library, named \textit{Default}. In this mechanism, for the handover to occur, the user must surpass the hysteresis point. This point corresponds to when the signal strength of the serving cell, combined with the hysteresis value, becomes lower than the signal strength of the neighboring cell (Event A3).

\subsection{Vehicular Network}

For the characterization of the vehicular network, we used real-world traces provided in \cite{b11}. The vehicles maintained an average speed of 14 to 16 meters per second over a distance of approximately 3 kilometers. The network deployed to cover this area comprised three gNodeBs operating in standalone (SA) mode on the n78 band.

\subsection{Results}

For the performance analysis, we used two applications: a video streaming based on edge computing and an OTA software update. This subsection details the configurations used to characterize these use cases and the results obtained.

\subsubsection{Video Streaming}

Video streaming has been gaining traction in vehicular networks, supporting computer vision and assisting sensors \cite{stream}, thus, this is the first use case to be analyzed. The video streaming used the Multi-Access Edge Computing (MEC) architecture provided by Simu5G through an application called \textit{RealTimeVideoStreamingApp}. In this setup, the MEC application (MEC APP) receives the video transmission sent by the vehicle via the \textit{MecRTVideoStreamingReceiver}, which manages the data sending. Simu5G provides traces that vary the data transmission over time.

For this use case, we analyzed the CQI, delay, and Session Freeze, which refers to interruptions in video playback, as QoS metrics. Finally, we conducted a statistical analysis using the ANOVA test to assess significant differences between the tested models. 

\begin{figure}[ht!]
    \centering
    \includegraphics[width=0.72\linewidth]{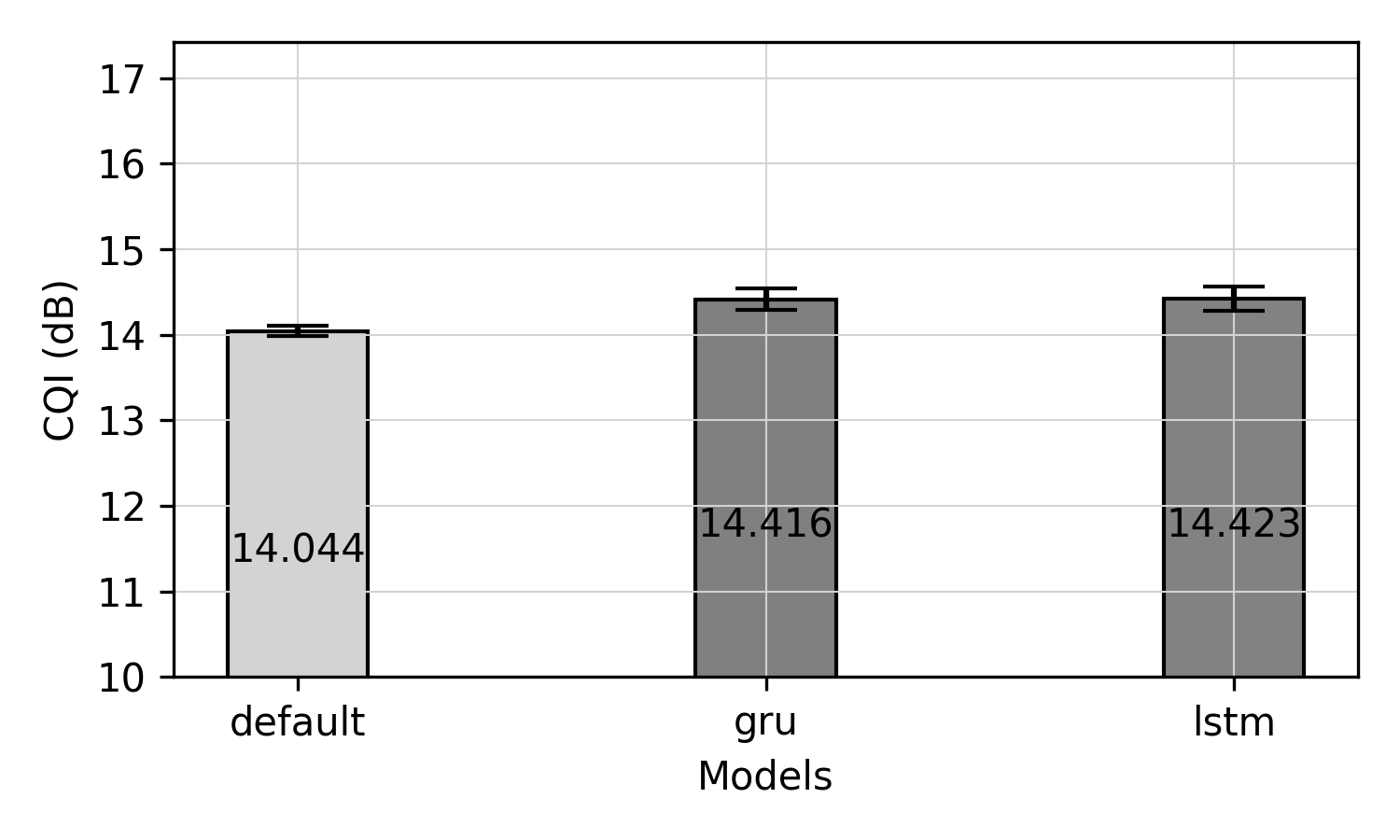}
    \caption{Channel Quality Indicator by Model.}
    \label{fig:cqi_stream}
\end{figure}

The first metric analyzed was the CQI, presented on the Figure \ref{fig:cqi_stream}. For this metric, the model that exhibited the best average was the LSTM, increasing by 0.049\% over GRU and 2.695\% over the default.

Using ANOVA, a p-value (p-val) less than 0.05 indicates that the results are statistically significant, suggesting that the data are likely different among the groups. Additionally, an F-statistic (F-val) greater than 1 indicates that the variability between group means is greater than the variability within the groups. As shown in Table \ref{tab:resultados_anova_stream}, LSTM and GRU differ from the default but show similar behavior when compared to each other. Both models demonstrate an improvement in user signal quality, attributed to the proactive anticipation of handover procedure. This enhancement is made possible by the real-time monitoring capabilities provided by Open RAN.

\begin{table}[h!]
    \centering
    \caption{Comparison based on ANOVA on Streaming User Case}
    \begin{tabular}{@{}lcccccccc@{}}
        \toprule
        & \multicolumn{2}{c}{CQI} & \multicolumn{2}{c}{Delay} & \multicolumn{2}{c}{Session Freeze} \\ 
        \cmidrule(lr){2-3} \cmidrule(lr){4-5} \cmidrule(lr){6-7}
        Comparison & F-val & p-val & F-val & p-val & F-val & p-val \\ 
        \midrule
        Default vs LSTM & 23.26 & 1e-05 & 34.42 & 2e-07  & 15.63 & 0.0002 \\ 
        Default vs GRU  & 25.96 & 4e-06 & 19.63 & 4e-05 & 10.86 & 0.001\\ 
        GRU vs LSTM     & 0.005  & 0.94  & 0.695 & 0.407 & 2.02 & 0.16 \\ 
        \bottomrule
    \end{tabular}
    \label{tab:resultados_anova_stream}
\end{table}

\begin{figure}[ht!]
    \centering
    \includegraphics[width=0.72\linewidth]{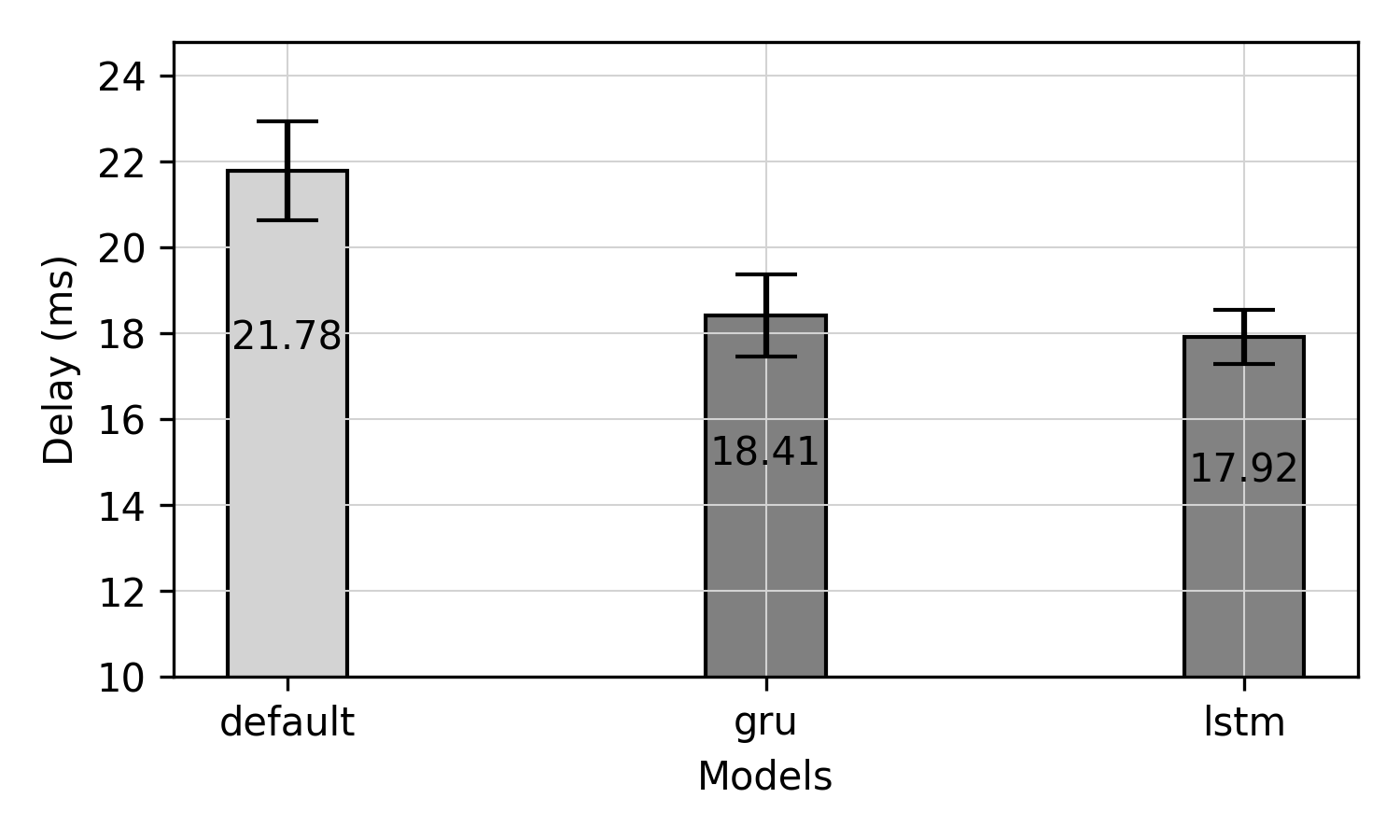}
    \caption{Delay by Model.}
    \label{fig:delay_stream}
\end{figure}

The first QoS metric analyzed to assess the impacts of the handover solutions was delay, presented on the Figure \ref{fig:delay_stream}. For this metric, the model that exhibited the best average was the LSTM, reducing the delay by approximately 2.67\% compared to the GRU and by about 17.71\% compared to the default. 

\begin{figure}[ht!]
    \centering
    \includegraphics[width=0.72\linewidth]{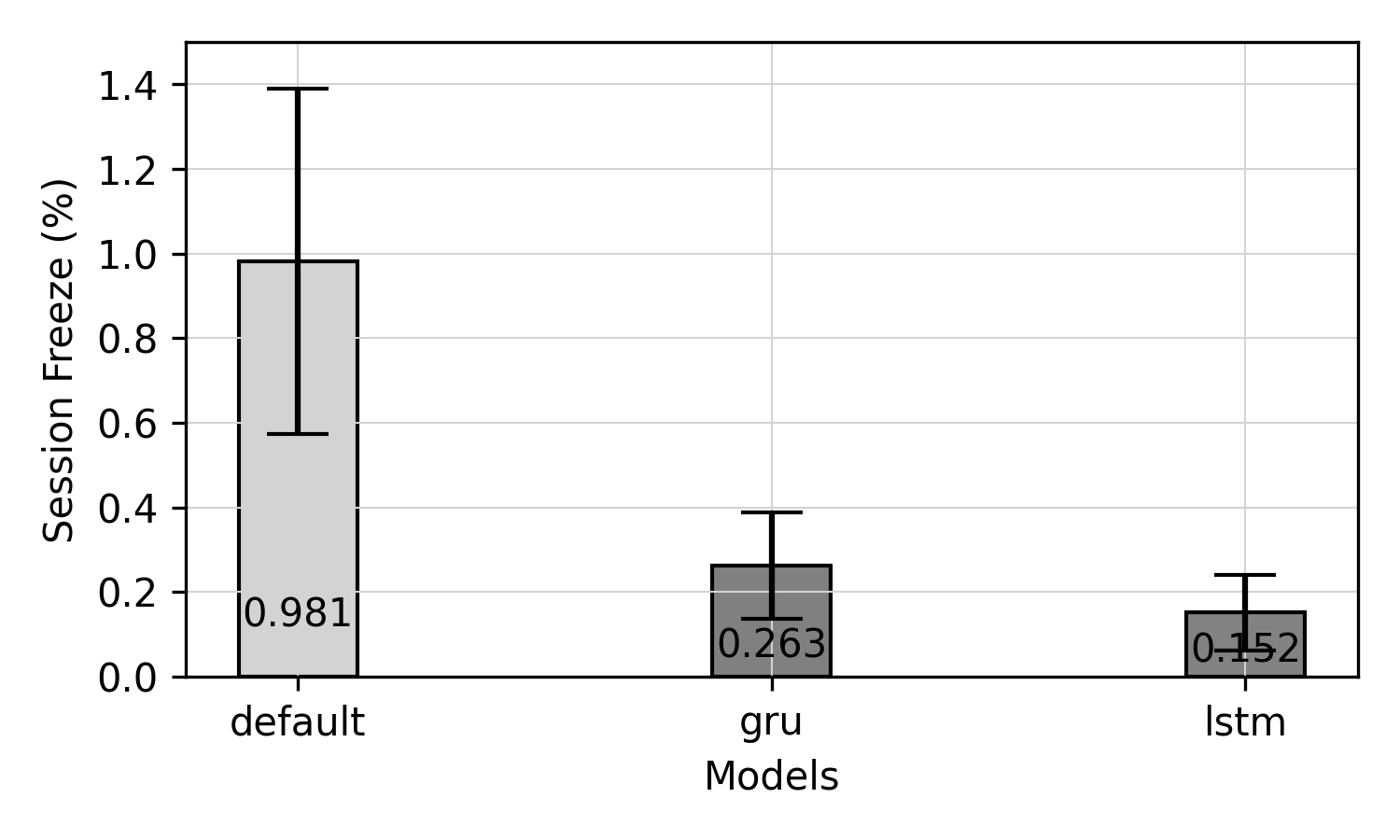}
    \caption{Session Freeze by Model.}
    \label{fig:freeze}
\end{figure}

The delay results reinforce the impacts on CQI, thus proving the effectiveness of the proposed solution. Finally, the last metric analyzed for this use case was session freeze, presented on the Figure \ref{fig:freeze}. Similar to the previous metrics, the LSTM and GRU exhibited statistically similar behaviors, again optimizing when compared to the default model.

\subsubsection{Over-the-Air (OTA) Software Updates}

OTA software updates are one of the emerging use cases in vehicular networks. This type of application has become widely feasible with the advent of 5G, which enables high data transmission rates \cite{b12}. This is the second use case evaluated in this article, which explores how handover optimization based on Open RAN can enhance the implementation of OTA software updates in vehicles, avoiding the interruption during the movement. For this scenario, we considered a software size of 35.6 MB, as presented in \cite{b12}. Its transmission was a Constant Bit Rate (CBR) application, sending packets of 1024 bytes every 5 ms. For this use case, the network metrics analyzed were the CQI and throughput.

\begin{figure}[ht!]
    \centering
    \includegraphics[width=0.72\linewidth]{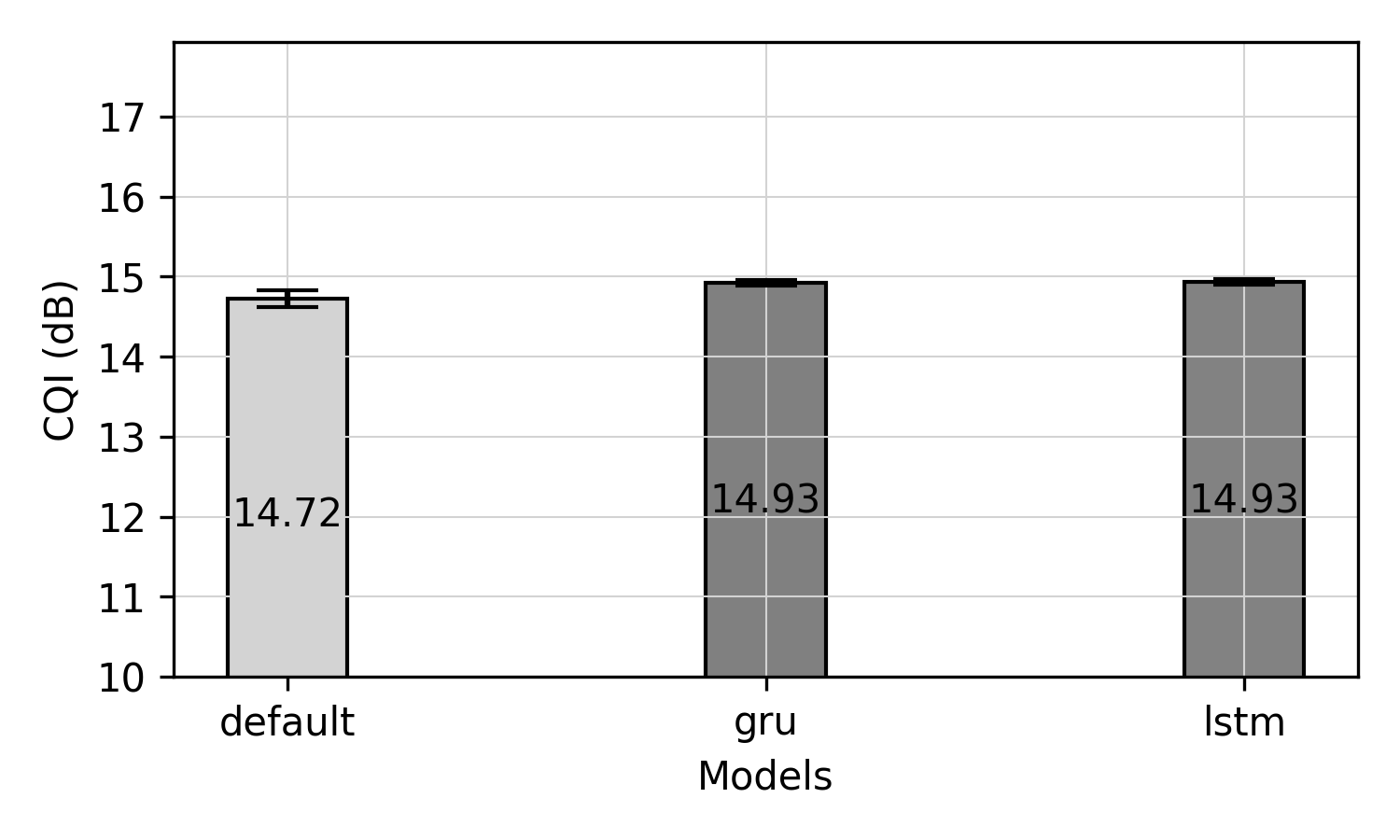}
    \caption{Channel Quality Indicator by Model.}
    \label{fig:cqi_ota}
\end{figure}

\begin{figure}[ht!]
    \centering
    \includegraphics[width=0.72\linewidth]{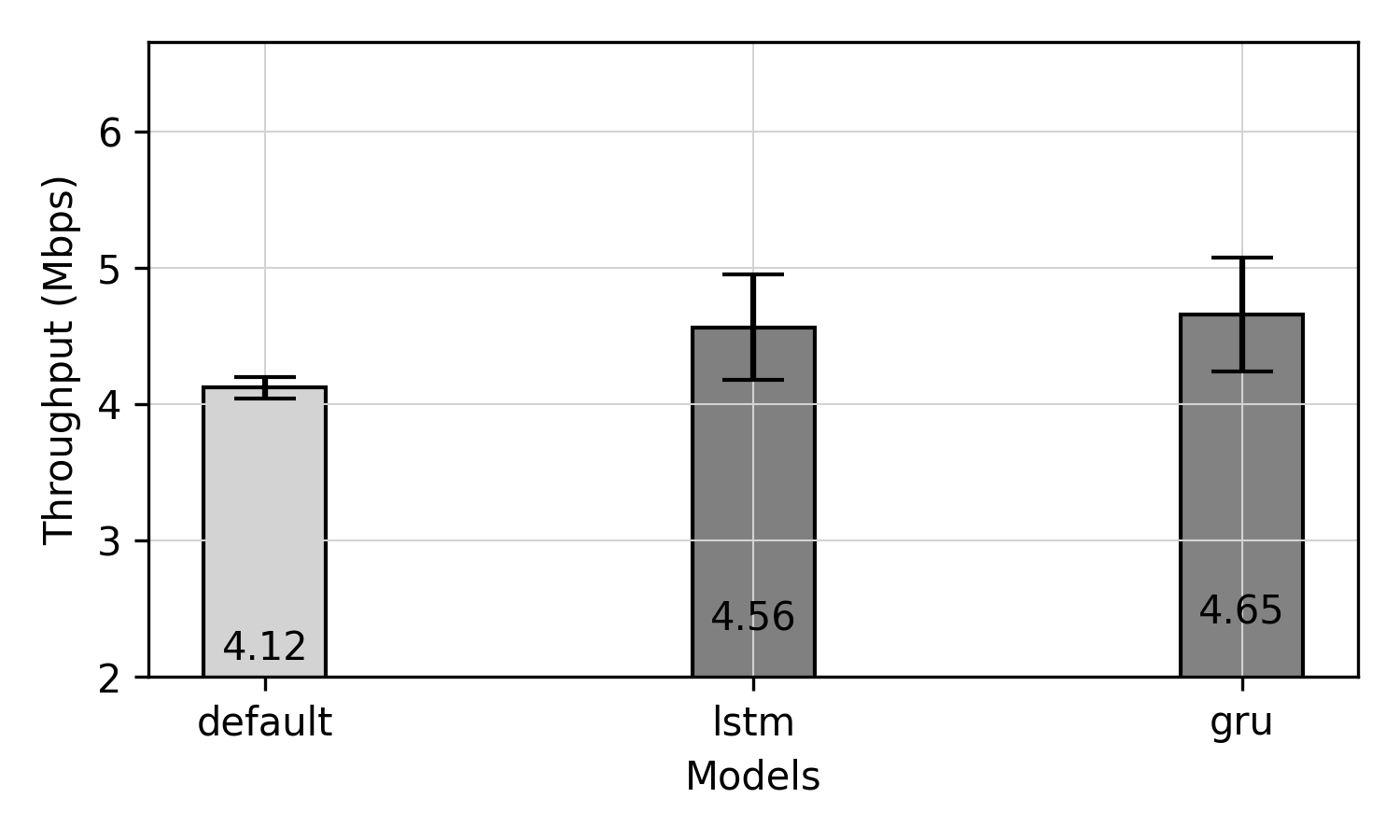}
    \caption{Mean Throughput by Model.}
    \label{fig:Throughput}
\end{figure}

For the CQI, presented in Figure \ref{fig:cqi_ota}, the models that exhibited the best average were the LSTM and GRU, increasing by approximately 1.42\% compared to the default. As can also be seen in Table \ref{tab:anova_ota}, the GRU and LSTM models showed statistically similar behaviors, differing only from the default model. The impacts of handover optimization can also be seen in the throughput, presented on the Figure \ref{fig:Throughput}, where the LSTM exhibited the best average value, with an increase of 1.97\% compared to the GRU and 12.85\% compared to the default.

\begin{table}[ht!]
    \centering
    \caption{Comparison based on ANOVA on OTA User Case}
    \begin{tabular}{@{}lcccccc@{}}
        \toprule
        & \multicolumn{2}{c}{CQI} & \multicolumn{2}{c}{Throughput} \\ 
        \cmidrule(lr){2-3} \cmidrule(lr){4-5}
        Comparison & F-val & p-val & F-val & p-val \\ 
        \midrule
        Default vs LSTM & 14.38 &  0.0004 & 4.80 & 0.03 \\ 
        Default vs GRU &  13.57 & 0.0006 & 6.03 &  0.01 \\ 
        GRU vs LSTM & 0.04 & 0.83 & 0.09 & 0.75 \\ 
        \bottomrule
    \end{tabular}
    \label{tab:anova_ota}
\end{table}

With these results, it can be observed that the proposed solution offers an enhancement in QoS metrics, due to the proactive handover anticipation process, thus demonstrating the efficiency of the solution.

\section{Conclusion}
\label{sec:conclusion}

This work proposed a handover mechanism based on deep learning using GRU and LSTM algorithms. We validated the model in a simulation environment in OMNeT++, integrating it with a containerized Near-RT RIC. The results indicated that both LSTM and GRU models effectively manage handover, showing similar performance in ANOVA tests. The LSTM model achieved a 2.7\% increase in CQI for video streaming and a 1.42\% increase for OTA scenarios compared to the standard model. The LSTM also reduced latency by approximately 17\% and increased throughput by 12.85\%. This demonstrates the effectiveness of the proposed solution, leveraging Open RAN and deep learning. Additionally, Open RAN allows for automated decision-making without requiring user-implemented monitoring mechanisms, as measurement report metrics serve as inputs for xApps. Future work will focus on utilizing a Traffic Steering xApp to alleviate load on gNodeBs and exploring emerging use cases for autonomous vehicles.

\section*{Acknowledgment}
This work was supported by the National Council for Scientific and Technological Development (CNPq) - Research Productivity Fellowship (Grant No. 313083/2023-1) and Pernambuco Research Foundation (FACEPE) (Grant No. IBPG-0130-1.03/23).

\end{document}